\documentclass[english,aps,preprint]{revtex4}
\usepackage[T1]{fontenc}
\usepackage[utf8]{luainputenc}
\usepackage{float}
\usepackage{amsmath}
\usepackage{amssymb}
\usepackage{graphicx}
\usepackage{setspace}
\usepackage{esint}

\makeatletter

\newcommand{\noun}[1]{\textsc{#1}}
\providecommand{\tabularnewline}{\\}

\@ifundefined{textcolor}{}
{%
 \definecolor{BLACK}{gray}{0}
 \definecolor{WHITE}{gray}{1}
 \definecolor{RED}{rgb}{1,0,0}
 \definecolor{GREEN}{rgb}{0,1,0}
 \definecolor{BLUE}{rgb}{0,0,1}
 \definecolor{CYAN}{cmyk}{1,0,0,0}
 \definecolor{MAGENTA}{cmyk}{0,1,0,0}
 \definecolor{YELLOW}{cmyk}{0,0,1,0}
}

\makeatother

\usepackage{babel}
\begin{document}

\title{Direct detection and solar capture of spin-dependent dark matter }

\author{Zheng-Liang Liang}

\email{liangzl@itp.ac.cn}

\author{Yue-Liang Wu}

\email{ylwu@itp.ac.cn}

\address{State Key Laboratory of Theoretical Physics (SKLTP),}

\address{Kavli Institute for Theoretical Physics China (KITPC),}

\address{Institute of Theoretical Physics, Chinese Academy of Science, Beijing
100190, China}

\address{University of Chinese Academy of Sciences, Beijing 100049, China}
\begin{abstract}
We investigate the implication of different elastic spin-dependent (SD) operators
on both the direct and indirect detections of the weakly interacting
massive particle (WIMP). Six representative building blocks of SD
operators, together with their counterparts with a massless mediator,
are considered to interpret the direct detection experiments~(Xenon100,
SIMPLE, and COUPP) in a comprehensive way. We also study the solar
capture and annihilation of WIMPs with these effective SD operators
and place the constraints on the relevant annihilation rate from neutrino
detection experiments Super-Kamionkande and IceCube. Upper limits
on the WIMP-nucleon couplings drawn from direct detections are also
projected to the annihilation rate for contrast and complementarity.
We find that the efficiency of these mentioned detection strategies
depends specifically on the six SD operators, while the neutrino-based
detections are more effective in exploring the parameter space for
the massless mediator scenario.
\end{abstract}
\maketitle
The existence of dark matter (DM) has  been well confirmed through
decades of endeavor in cosmological observation~\cite{Jungman:1995df,Bertone:2004pz}.
However, the nature of the DM still remains a challenging problem
for particle physics. Among those mechanisms and DM candidates, the
weakly interacting massive particle (WIMP) is one of the most promising.  In this picture, the WIMPs interact with the standard model particles
through weak interactions and naturally result in the thermal relic
density consistent with the observation, called the ``WIMP
miracle.'' Both theoretical and experimental attempts have been made
to account for and search for such particles. Many direct detection
groups have reported their results based on the conventional elastic
spin-independent (SI) or spin-dependent (SD) effective operator that
is proportional to a constant, $\mathcal{M}_{\mathrm{SI}}\propto1$,
or to the dot product between the spins of the WIMP and quark, $\mathcal{M}_{\mathrm{SD}}\propto\mathbf{S}_{\chi}\cdot\mathbf{S}_{q}$.
However, if we broaden our study to more general DM scenarios such
as isospin-violating DM~\cite{Feng:2011vu}, long-range force DM~\cite{Schwetz:2011xm,Kumar:2012uh},
inelastic DM~\cite{TuckerSmith:2001hy}, and form factor DM~\cite{Feldstein:2009tr},
the interpretation of the experimental results may turn out to be
quite different. This possibility has been used in attempts to alleviate
the conflicts between different direct detection experiments~\cite{Masso:2009mu,Barger:2010gv,Chang:2010en,Fitzpatrick:2010br,Farina:2011pw,Fornengo:2011sz,Foot:2011pi,Frandsen:2011ts,DelNobile:2012tx,Haisch:2013uaa,Frandsen:2013cna}.

In recent years, several groups began to systematically study the
DM direct detection in terms of the nonrelativistic (NR) effective
field theory~\cite{Fan:2010gt,Fitzpatrick:2012ix,Fitzpatrick:2012ib,DelNobile:2013sia},
in which a given Lorentz-invariant interaction is expanded with a
complete set of NR effective operators at the nucleon level, along with
the corresponding Galilean and rotational invariant coefficients.
The authors of Ref. \cite{Fitzpatrick:2012ix} have further calculated the relevant
nuclear form factors associated with different effective operators
beyond the simplest SI and SD case, for the elements common in the
present-day detectors. Besides, a state-of-the-art large-scale nuclear
structure calculation for different operators has also been planned
in \cite{Klos:2013rwa}. Further knowledge about the nuclear form factors
is important for us to interpret the detection results in a more comprehensive
way.

Besides direct detection efforts, some indirect approaches are also
expected to be effective in constraining the coupling strength between
the WIMP and nucleon~\cite{Belotsky:2002sv,Hooper:2008cf,Halzen:2009vu,Kappl:2011kz}.
One may regard the solar neutrinos that are detected by the neutrino
detectors (e.g., Super-Kamionkande~\cite{Tanaka:2011uf} or IceCube~\cite{Abbasi2012,Aartsen:2012kia})
as the possible annihilation products of the trapped WIMPs residing
in the center region of the Sun, so as to impose upper limits on the
WIMP-nucleus coupling coefficients, which is relevant to the WIMP capture
rate and thus to the neutrino flux, if an equilibrium between capture
and annihilation is assumed.

The implications of a wide variety of elastic SI interactions on both the
direct and indirect detection experiments have been studied in Ref.~\cite{Guo:2013ypa},
in which the direct detections are proved to be more effective in
constraining the WIMP-nucleon interaction couplings with a sensitivity
$2\sim4$ orders of magnitude greater than the neutrino-based approach.
However, given that the upper limit on the elastic SD WIMP-nucleon cross section
is far above that of the SI interaction, one would expect the neutrino-based
detection to compete with or even exceed the direct detections in
sensitivity for effective SD operators. We will study in this paper
the direct detection of the WIMPs with a set of effective elastic SD operators,
as well as the relevant indirect detection in a similar way with \cite{Guo:2013ypa}
and compare them with each other in sensitivity. We organize these
discussions as follows: In Sec. \uppercase\expandafter{\romannumeral1} we discuss the effective SD
operators in a systematic manner and the relevant consequences for
various WIMP direct detection experiments. In Sec. \uppercase\expandafter{\romannumeral2} we study
the solar capture of the WIMP in detail and calculate the bounds on the capture
rate $C_{\odot}$, and the annihilation rate $\Gamma_{\odot}=C_{\odot}/2$,
imposed by both the direct detection experiments and the neutrino
detectors Super-Kamionkande and IceCube, for the purpose of comparison
and complementarity. Conclusions and discussions are given in Sec.
\uppercase\expandafter{\romannumeral3}.

\section{dark matter direct detection}

\subsection{Direct detection recoil rate}

In general the differential event rate $R$ in a direct detection
experiment is given as an average over the WIMP distribution,

\begin{equation}
\frac{dR}{dE_{R}}=N_{T}\frac{\rho_{\chi}}{m_{\chi}}\int_{v_{\mathrm{min}}}^{v_{\mathrm{e}}}\frac{d\sigma}{dE_{R}}vf(\mathbf{v})d^{3}v,\label{eq:differential event rate1}
\end{equation}
where $N_{T}$ is the effective number of target nucleus in the detector,
and $\rho_{\chi}$ is the local WIMP halo density in our earth neighborhood, with
$m_{\chi}$ being the WIMP mass. The WIMP velocity distribution $f(\mathbf{v})$
is defined in the laboratory reference frame with the incident WIMP
velocity $\mathbf{v}$, and $d\sigma/dE_{R}$ is the relevant WIMP-nucleus
differential cross section, which can be further expressed (by summing
over initial spins and averaging over finial spins) as

\begin{equation}
\frac{d\sigma}{dE_{R}}=\frac{m_{T}}{2\pi v^{2}}\frac{1}{(2J+1)(2s_{\chi}+1)}\sum_{\mathrm{spins}}|\mathcal{M}_{\mathrm{NR}}|^{2}.\label{eq:differential cross section}
\end{equation}

We denote the nucleus spin and WIMP spin as $J$ and $s_{\chi}$, respectively,
and $m_{T}$ is the mass of the target nucleus. $\mathcal{M}_{NR}$
is the nonrelativistic scattering amplitude, which differs from the
relativistic one $\mathcal{M}$ by $\mathcal{M}_{NR}=\mathcal{M}/(4m_{\chi}m_{T})$.
Thus by Eq.~(\ref{eq:differential cross section}), Eq.~(\ref{eq:differential event rate1})
can be further written as

\begin{equation}
\frac{dR}{dE_{R}}=N_{T}\frac{\rho_{\chi}}{m_{\chi}}\int_{v_{\mathrm{min}}}^{v_{\mathrm{e}}}\frac{m_{T}}{2\pi}\frac{1}{(2J+1)(2s_{\chi}+1)}\sum_{\mathrm{spins}}|\mathcal{M}_{\mathrm{NR}}|^{2}\frac{f(\mathbf{v})}{v}d^{3}v.
\end{equation}

The upper limit of integral $v_{e}$ is the galactic escape velocity
relative to the detector, and the lower limit $v_{\mathrm{min}}$
is the minimal possible velocity for a fixed recoil energy $E_{R}$,
which is related to the nucleus mass $m_{T}$ and reduced mass $\mu_{T}$
of the WIMP-nucleus pair through
\begin{equation}
v_{\mathrm{min}}=\frac{q}{2\mu_{T}},\label{eq:vmin}
\end{equation}
where $q=\sqrt{2m_{T}E_{R}}$ is the transferred momentum for elastic scattering.

\subsection{Elastic spin-dependent dark matter form factors}

Instead of listing possible Lorentz-invariant effective operators
in our model-independent analysis, we take an alternative strategy
by following~\cite{Anand:2013yka} to enumerate a set of effective
operators at the nonrelativistic level, which preserve Galilean invariance
and rotational symmetry degenerated from Lorentz symmetry, as well
as their corresponding discrete symmetry at the relativistic level.
These independent 15 operators with their invariant coefficients exhausting
all possible nonrelativistic effective operators for spin 1/2 WIMP,
which originate from the nonrelativistic reduction of all possible
20 bilinear amplitude products~\cite{Fitzpatrick:2012ix,Anand:2013yka}, may also be encountered in higher-spin WIMP models.
We first divide the following 11 effective WIMP-nucleon operators
into the following six groups separately in six lines:

1. P-even, $\mathbf{S}_{\chi}$-independent,  T-even
\[
\mathcal{O}_{1,N}=1,\quad\mathcal{O}_{2,N}=(v^{\perp})^{2},\quad\mathcal{O}_{3,N}=i\mathbf{S}_{N}\cdot(\mathbf{q}\times\mathbf{v}^{\perp}),
\]

2. P-even, $\mathbf{S}_{\chi}$-dependent, T-even
\[
\mathcal{O}_{4,N}=\mathbf{S}_{\chi}\cdot\mathbf{S}_{N},\quad\mathcal{O}_{5,N}=i\mathbf{S}_{\chi}\cdot(\mathbf{q}\times\mathbf{v}^{\perp}),\quad\mathcal{O}_{6,N}=(\mathbf{S}_{\chi}\cdot\mathbf{q})(\mathbf{S}_{N}\cdot\mathbf{q}),
\]

3. P-odd, $\mathbf{S}_{\chi}$-independent, T-even
\[
\mathcal{O}_{7,N}=\mathbf{S}_{N}\cdot\mathbf{v}^{\perp},
\]

4. P-odd, $\mathbf{S}_{\chi}$-dependent, T-even
\[
\mathcal{O}_{8,N}=\mathbf{S}_{\chi}\cdot\mathbf{v}^{\perp},\quad\mathcal{O}_{9,N}=i\mathbf{S}_{\chi}\cdot(\mathbf{S}_{N}\times\mathbf{q}),
\]

5. P-odd, $\mathbf{S}_{\chi}$-independent, T-odd
\[
\mathcal{O}_{10,N}=i\mathbf{S}_{N}\cdot\mathbf{q},
\]

6. P-odd, $\mathbf{S}_{\chi}$-dependent, T-odd
\begin{equation}
\mathcal{O}_{11,N}=i\mathbf{S}_{\chi}\cdot\mathbf{q}, \label{eq:operators}
\end{equation}
where $\mathbf{q}$ is the momentum transferred to the WIMP.  It is easy to verify that each operator from one group does not interfere with operators from another
 by the consideration of symmetry and the WIMP spin $\mathbf{S}_{\chi}$. In addition we list another three operators that are constructed
from the mixing  among the above operators,

\begin{equation}
\mathcal{O}_{10,N}\mathcal{O}_{8,N},\;\mathcal{O}_{11,N}\mathcal{O}_{7,N},\;\mathcal{O}_{11,N}\mathcal{O}_{3,N},
\end{equation}
and one last building block to form a complete set,

\begin{equation}
\mathcal{O}_{12,N}=\mathbf{S}_{\chi}\cdot(\mathbf{S}_{N}\times\mathbf{v}^{\perp}).
\end{equation}

The Hermitian ``perpendicular'' operator $\mathbf{v}^{\perp}=\mathbf{v}+\frac{\mathbf{q}}{2\mu_{N}}$
satisfies $\mathbf{v}^{\perp}\cdot\mathbf{q}=0$ when the on-shell
condition is imposed, and $\mathbf{v}$ is the velocity of the WIMP
with respect to the nucleon. One should note that only SD operators
$\mathcal{O}_{3,N},$ $\mathcal{O}_{4,N}$, $\mathcal{O}_{6,N}$,
$\mathcal{O}_{7,N}$, $\mathcal{O}_{9,N}$, and $\mathcal{O}_{10,N}$
are included in our study in this work, because these element operators
are not only frequently encountered in practical operator expansion
but also sufficient to give us a general description on how a variety
of SD effective operators with different powers of $q$ and $v^{\perp}$
will lead to different interpretations of DM detection experiments.
By the notation of \cite{Fitzpatrick:2012ix} , the differential recoil
rate for the conventional SD operator $\mathcal{O}=a_{p}\mathbf{S}_{\chi}\cdot\mathbf{S}_{p}$
is written as

\begin{equation}
\frac{dR}{dq}=N_{T}\frac{\rho_{\chi}}{m_{\chi}}\frac{16}{3}\frac{\sigma_{p}}{2\mu_{p}^{2}}\frac{(F_{\Sigma^{''}}^{(p,p)}+F_{\Sigma^{'}}^{(p,p)})}{16}q\int_{v_{\mathrm{min}}}^{v_{\mathrm{e}}}\frac{f(\mathbf{v})}{v}d^{3}v,\label{eq:differential even rate3}
\end{equation}
and the WIMP-proton cross section
$\sigma_{p}$ and coupling $a_{p}$ are connected to each other by

\begin{equation}
\frac{\sigma_{p}}{2\mu_{p}^{2}}=C(s_{\chi})\frac{a_{p}^{2}}{2\pi}\frac{3}{16},\label{eq:cross section-coupling relation}
\end{equation}
where $F_{\Sigma^{'}}^{(p,p)}$ and $F_{\Sigma^{''}}^{(p,p)}$ are the transverse and the longitudinal form factors respectively~\cite{Fitzpatrick:2012ix}, and $C(s_{\chi})=\frac{4}{3}s_{\chi}(s_{\chi}+1)$ is normalized
to unit for a Dirac fermion WIMP. By introducing the WIMP-nucleus form factor $F_{\chi-T}^{2}$, other
SD operators can be generalized and expressed in a consistent form
with Eq.~(\ref{eq:differential even rate3}) as

\begin{equation}
\frac{dR}{dq}=N_{T}\frac{\rho_{\chi}}{m_{\chi}}\frac{16}{3}\frac{\sigma}{2\mu_{p}^{2}}q\int_{v_{\mathrm{min}}}^{v_{e}}F_{\chi-T}^{2}(q/q_{0},\, v/v_{0},\, a_{p}/a_{n})\frac{f(\mathbf{v})}{v}d^{3}v.\label{eq:differential event rate4}
\end{equation}

Two reference parameters $q_{0}=100~ \mathrm{MeV}$, the typical scale of
the momentum transferred in common direct detection experiments, and $v_{0}=220~\mathrm{km/s}$,
the WIMP velocity dispersion, are brought in
to keep $F_{\chi-T}^{2}$ a dimensionless factor. The nominal
``cross section'' $\sigma$ here is only a parameter that encodes
the coupling strength $a$. The mass difference between proton and
neutron is ignored in this work. In Table \ref{tab:dark-matter form factor},
we also normalize these operators to dimensionless ones with $q_{0}$
and $v_{0}$, in line with the conventional SD operator $\mathcal{O}_{4}$,
and absorb the relevant couplings \{$a_{N}$\} into parameter $\sigma$.
In this paper, we also consider a light mediator scenario in which
the WIMP-nucleon interaction is mediated by a low mass particle and
in the zero mass limit a $1/q^{2}$ factor arising from the massless
propagator appears as the coefficient of each operator in Eq.~(\ref{eq:operators}).
All these six SD operators and their counterparts with light mediator
are summarized in Table \ref{tab:dark-matter form factor}. For the
purpose of illustration we take $\mathcal{O}=\sum a_{N}\mathcal{O}_{9,N}=a_{p}i\mathbf{S}_{\chi}\cdot(\mathbf{S}_{p}\times\mathbf{q})+a_{n}i\mathbf{S}_{\chi}\cdot(\mathbf{S}_{n}\times\mathbf{q})$
as an example, with equal couplings $a=a_{p}=a_{n}$. From Table \ref{tab:dark-matter form factor}
we have:

\begin{equation}
F_{\chi-T}^{2}=\sum_{N,\, N'=p,n}C^{-1}(s_{\chi})F_{9,9}^{(N,\, N')}/q_{0}^{2},
\end{equation}
and

\begin{equation}
\frac{\sigma}{2\mu_{p}^{2}}=C(s_{\chi})\frac{a^{2}}{2\pi}\frac{3q_{0}^{2}}{16}.
\end{equation}

\{$F_{i,i}^{(N,\, N')}$\} are defined in Ref. \cite{Fitzpatrick:2012ix}.

\begin{table}
\begin{singlespace}
\begin{centering}
\begin{tabular}{|c|c|c|c|}
\hline
operator & normalized operator & parameter($\sigma$)  & $F_{\chi-T}^{2}$\tabularnewline
\hline
\hline
$\mathcal{O}_{3,N}$ & $\mathcal{O}_{3,N}/(q_{0}v_{0})$ & $f\, a^{2}q_{0}^{2}v_{0}^{2}$ & $\sum\frac{a_{N}a_{N'}}{a^{2}}F_{3,3}^{(N,\, N')}/(q_{0}^{2}v_{0}^{2})$\tabularnewline
\hline
$\mathcal{O}_{3,N}/q^{2}$ & $\mathcal{O}_{3,N}(q_{0}/q^{2}v_{0})$ & $f\,a^{2}v_{0}^{2}/q_{0}^{2}$ & $\sum\frac{a_{N}a_{N'}}{a^{2}}F_{3,3}^{(N,\, N')}(q_{0}^{2}/v_{0}^{2})$\tabularnewline
\hline
$\mathcal{O}_{4,N}$ & $\mathcal{O}_{4,N}$ & $f\,C(s_{\chi}) a^{2}$ & $\sum\frac{a_{N}a_{N'}}{a^{2}}C^{-1}(s_{\chi})F_{4,4}^{(N,\, N')}$\tabularnewline
\hline
$\mathcal{O}_{4,N}/q^{2}$ & $\mathcal{O}_{4,N}(q_{0}^{2}/q^{2})$ & $f\,C(s_{\chi})a^{2}/q_{0}^{2}$ & $\sum\frac{a_{N}a_{N'}}{a^{2}}C^{-1}(s_{\chi})F_{4,4}^{(N,\, N')}(q_{0}^{4}/q^{4})$\tabularnewline
\hline
$\mathcal{O}_{6,N}$ & $\mathcal{O}_{6,N}/q_{0}^{2}$ & $f\,C(s_{\chi}) a^{2}q_{0}^{4}$ & $\sum\frac{a_{N}a_{N'}}{a^{2}}C^{-1}(s_{\chi})F_{6,6}^{(N,\, N')}/q_{0}^{4}$\tabularnewline
\hline
$\mathcal{O}_{6,N}/q^{2}$ & $\mathcal{O}_{6,N}/q^{2}$ & $f\,C(s_{\chi}) a^{2}$ & $\sum\frac{a_{N}a_{N'}}{a^{2}}C^{-1}(s_{\chi})F_{6,6}^{(N,\, N')}/q^{4}$\tabularnewline
\hline
$\mathcal{O}_{7,N}$ & $\mathcal{O}_{7,N}/v_{0}$ & $f\,a^{2}v_{0}^{2}$ & $\sum\frac{a_{N}a_{N'}}{a^{2}}F_{7,7}^{(N,\, N')}/v_{0}^{2}$\tabularnewline
\hline
$\mathcal{O}_{7,N}/q^{2}$ & $\mathcal{O}_{7,N}q_{0}^{2}/(v_{0}q^{2})$ & $f\,a^{2}v_{0}^{2}/q_{0}^{4}$ & $\sum\frac{a_{N}a_{N'}}{a^{2}}F_{7,7}^{(N,\, N')}(q_{0}^{4}/v_{0}^{2}q^{4})$\tabularnewline
\hline
$\mathcal{O}_{9,N}$ & $\mathcal{O}_{9,N}/q_{0}$ & $f\,C(s_{\chi}) a^{2}q_{0}^{2}$ & $\sum\frac{a_{N}a_{N'}}{a^{2}}C^{-1}(s_{\chi})F_{9,9}^{(N,\, N')}/q_{0}^{2}$\tabularnewline
\hline
$\mathcal{O}_{9,N}/q^{2}$ & $\mathcal{O}_{9,N}(q_{0}/q^{2})$ & $f\,C(s_{\chi}) a^{2}/q_{0}^{2}$ & $\sum\frac{a_{N}a_{N'}}{a^{2}}C^{-1}(s_{\chi})F_{9,9}^{(N,\, N')}(q_{0}^{2}/q^{4})$\tabularnewline
\hline
$\mathcal{O}_{10,N}$ & $\mathcal{O}_{10,N}/q_{0}$ & $f\,a^{2}q_{0}^{2}$ & $\sum\frac{a_{N}a_{N'}}{a^{2}}4C^{-1}(s_{\chi})F_{6,6}^{(N,\, N')}/(q^{2}q_{0}^{2})$\tabularnewline
\hline
$\mathcal{O}_{10,N}/q^{2}$ & $\mathcal{O}_{10,N}(q_{0}/q^{2})$ & $f\, a^{2}/q_{0}^{2}$ & $\sum\frac{a_{N}a_{N'}}{a^{2}}4C^{-1}(s_{\chi})F_{6,6}^{(N,\, N')}(q_{0}^{2}/q^{6})$\tabularnewline
\hline
\end{tabular}
\par\end{centering}
\end{singlespace}

\caption{The SD operators $\{\mathcal{O}_{i}\}\,(i=3,\,4,\,6,\,7,\,9,\,10)$
together with corresponding amplitudes and form factors $F_{\chi-T}^{2}$.
\{$F_{i,i}^{(N,\, N')}$\} are defined in Ref.~\cite{Fitzpatrick:2012ix}.
The summations are taken over proton and neutron, $N,\, N'=p,\, n$.
Three choices of parameter $a=a_{p}(a_{n}=0),\, a=a_{n}(a_{p}=0)\,\mathrm{and}\: a=a_{p}=a_{n}(a_{p}=a_{n})$
correspond to the three situations explained in the text. The normalized
operators are listed in the second column and $f=\frac{3}{16}\frac{\mu_{p}^{2}}{\pi}$.
\label{tab:dark-matter form factor}}
\end{table}

\subsection{Direct detection constraints}

In this section, we try to study the constraints imposed by direct
detection experiments on couplings of the effective SD operators  in
Table \ref{tab:dark-matter form factor} for elastic scattering, where the corresponding form
factors are taken from Ref.~\cite{Fitzpatrick:2012ix}. In order to
interpret such constraints in a uniform way, we translate the experimental
bounds into the constraints on the effective cross section parameter
$\sigma$ through Eq.~(\ref{eq:differential event rate4}), and to
show the implications of ratio $a_{p}/a_{n}$ in our analysis, we
consider the following three extreme situations for reference: $a_{n}=0,$ $a_{p}=0,$
and $a_{p}=a_{n}$.

\begin{figure}[H]
\noindent \begin{centering}
\includegraphics[scale=0.45]{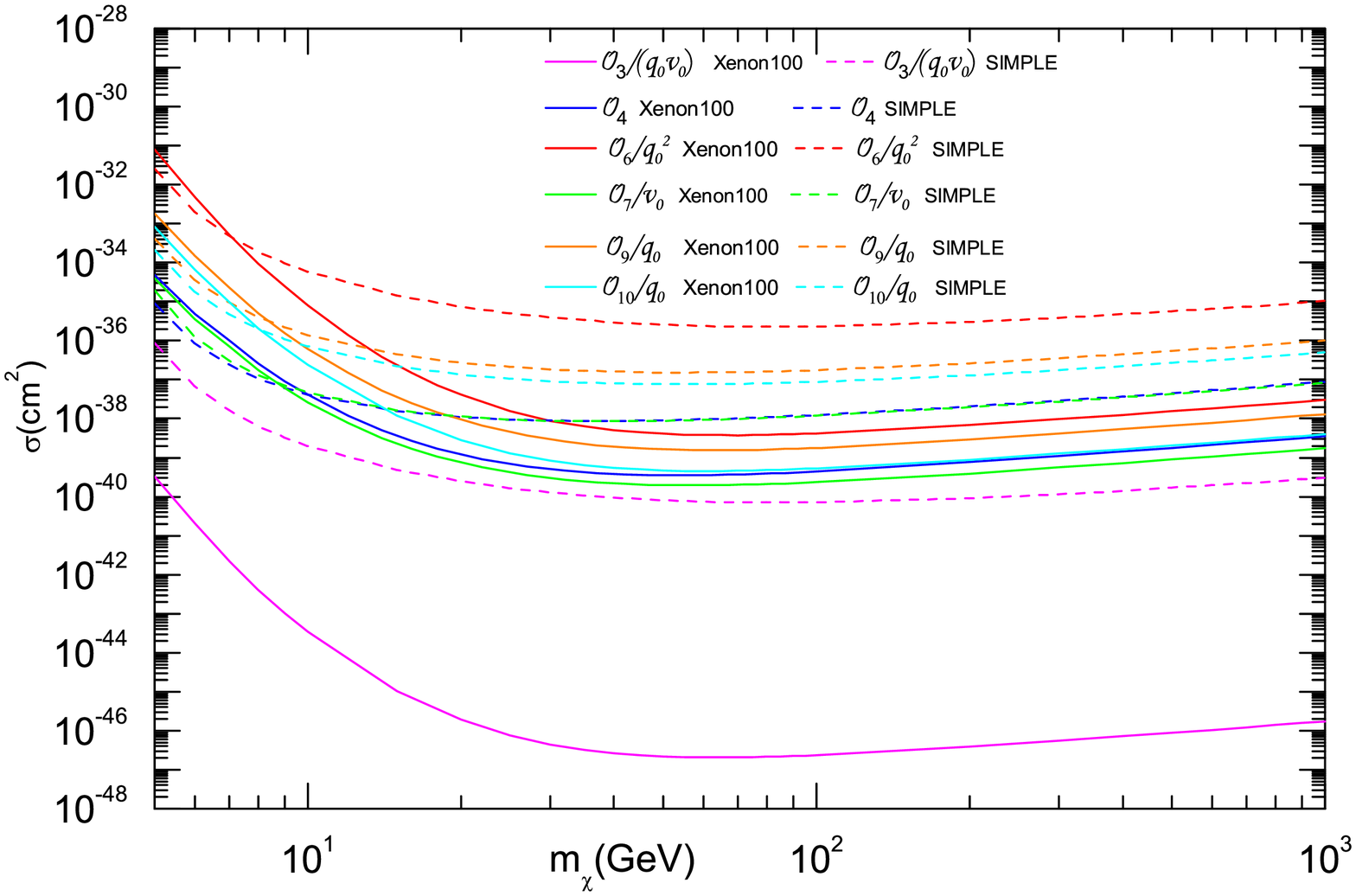}
\par\end{centering}

\noindent \begin{centering}
\includegraphics[scale=0.45]{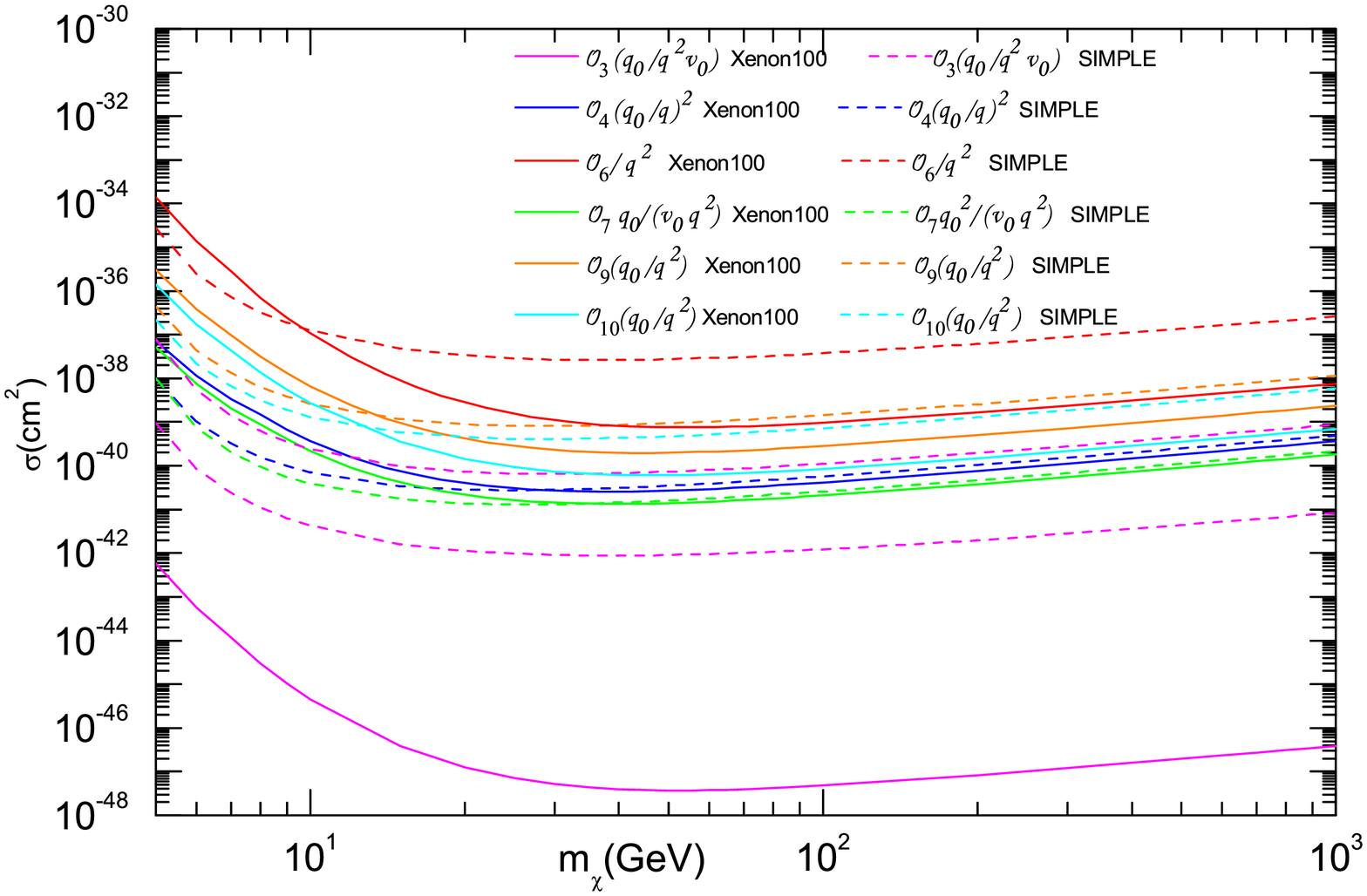}
\par\end{centering}

\caption{We combine the SIMPLE and Xenon100 90\% C.L. constraints on $\sigma$
for the normalized operators listed in Table \ref{tab:dark-matter form factor} for elastic scattering,
which are respectively the most sensitive in the low and high WIMP
mass ranges, shown separately in two panels for the purpose of clear
illustration, under the assumption $a_{p}=a_{n}$ .\label{fig:90=000025-C.L.-exclusion pn}}
\end{figure}

\begin{figure}[H]
\begin{centering}
\includegraphics[clip,scale=0.42]{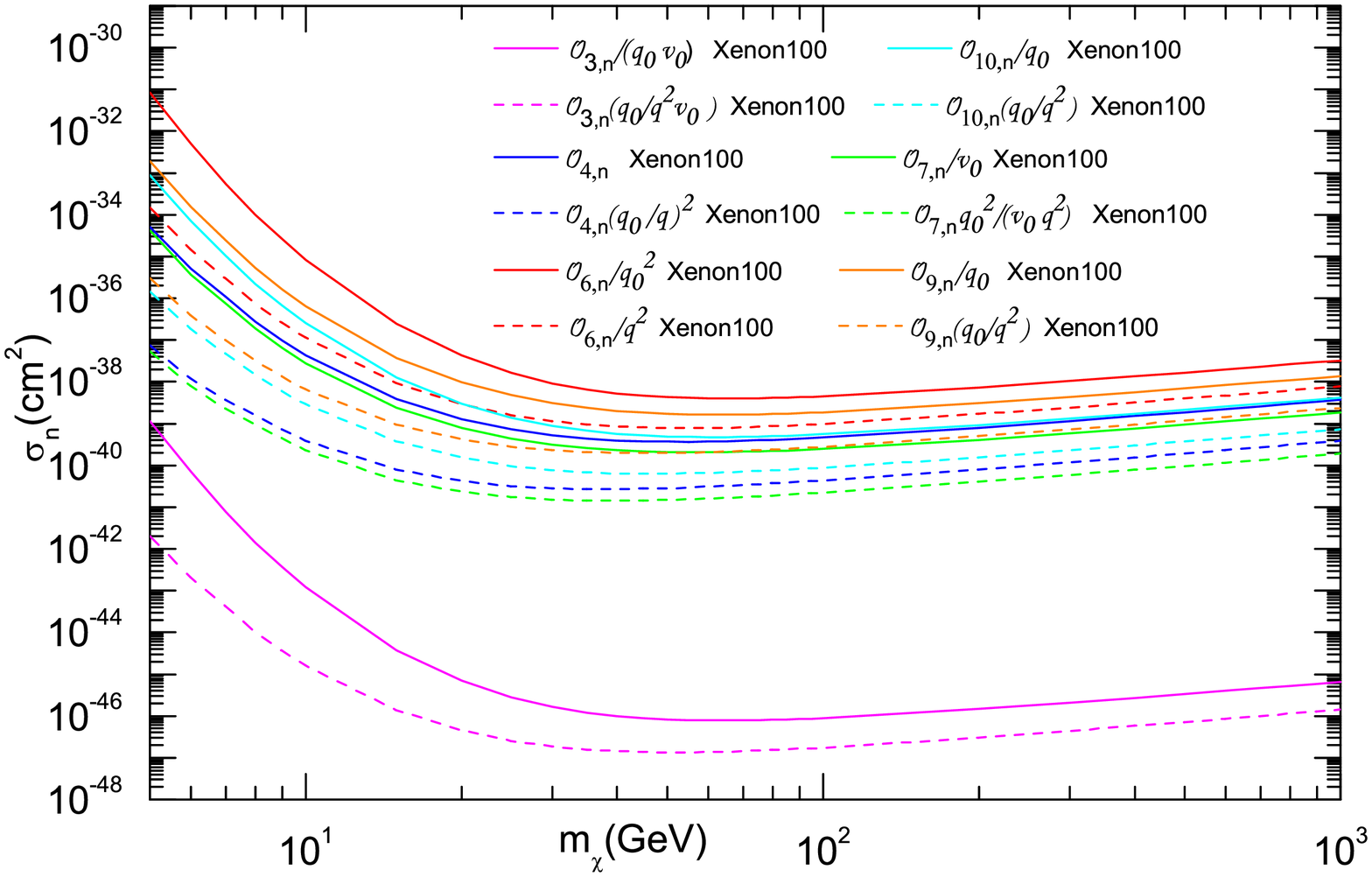}
\par\end{centering}

\begin{centering}
\includegraphics[clip,scale=0.42]{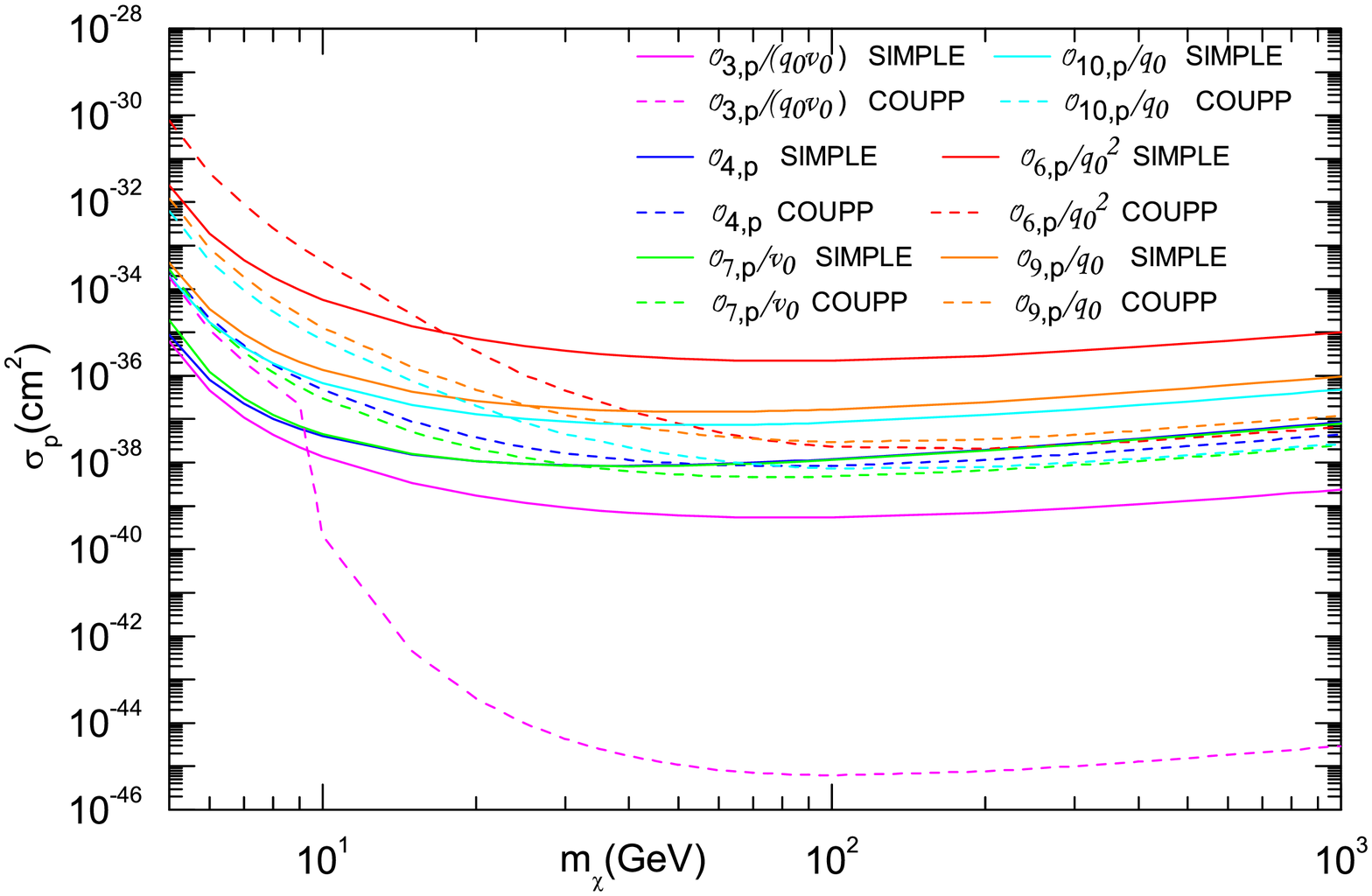} \includegraphics[clip,scale=0.42]{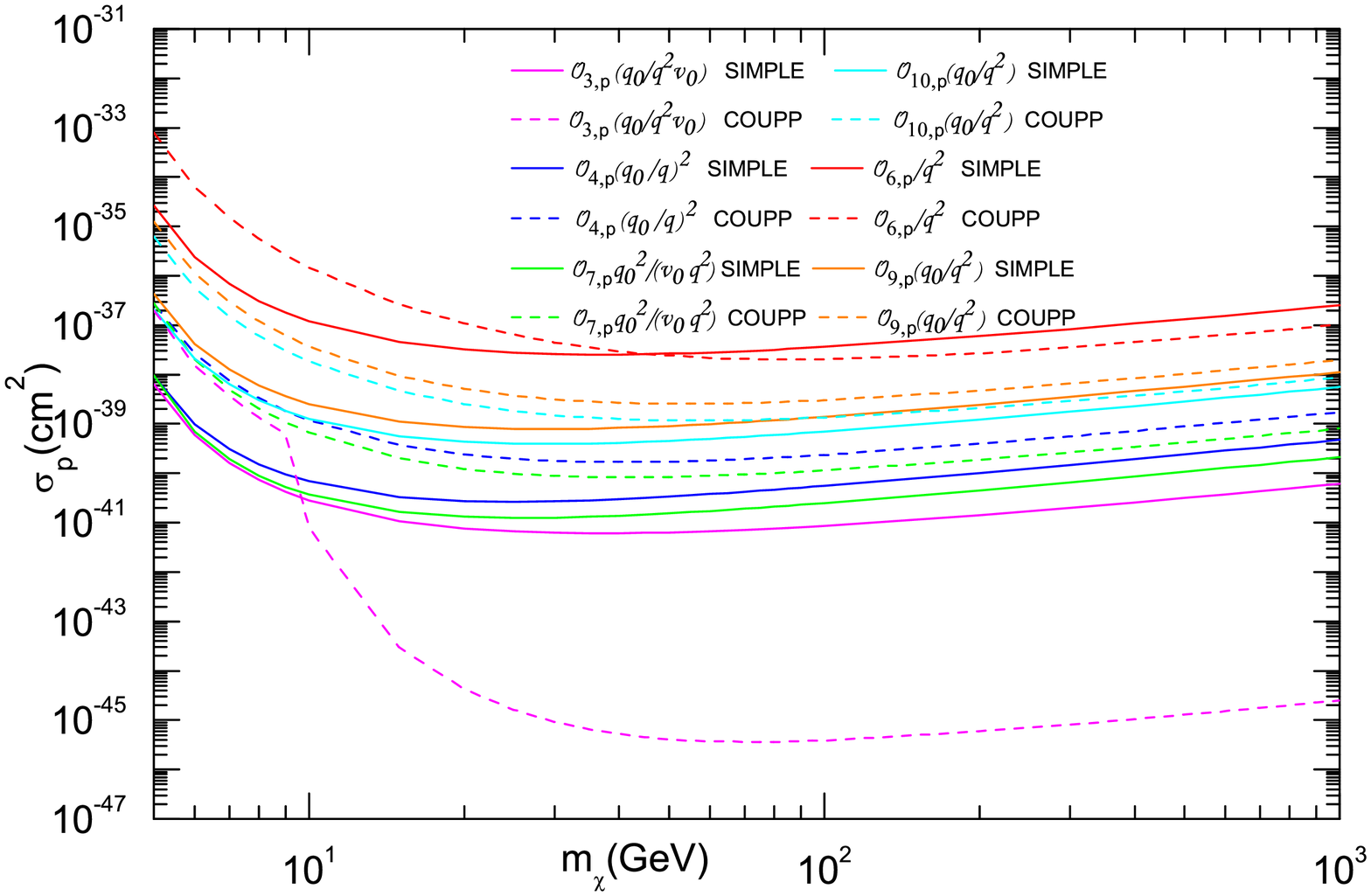}
\par\end{centering}

\caption{Xenon100 90\% C.L. upper limit on $\sigma_{n}$ for the normalized
operators in Table \ref{tab:dark-matter form factor} for elastic scattering assuming $a_{p}=0$(top
panel), and 90\% C.L. constraints on $\sigma_{p}$ from SIMPLE and
COUPP for the case $a_{n}=0$(middle and bottom panels) from Table
\ref{tab:dark-matter form factor}.\label{fig:90=000025-C.L.-exclusion n,p-1}}
\end{figure}

Due to the unpaired proton in fluorine, and iodine, SIMPLE and COUPP
are expected to possess a high efficiency in exploring the parameter
space for the case $a_{n}=0$. If the WIMP couples exclusively to the
neutron, we can use Xenon100 to place constraints by taking advantage
of the odd number of neutrons in Xenon isotopes. As for the case of
equal couplings $a_{p}=a_{n}$, comparisons between bounds from Xenon100
(neutron sensitive) and SIMPLE (proton sensitive) are also made in
Fig.~\ref{fig:90=000025-C.L.-exclusion pn}.

We use the reanalyzed
data from the first stage of the phase \uppercase\expandafter{\romannumeral2} SIMPLE dark matter search~\cite{Felizardo:2011uw},
of which 5 out of 14 previous candidate events are attributed to background,
reducing the expected signal rate to 0.289/events/kgd at 90\% C.L.
A more elaborate bubble nucleation efficiency $\eta=1-\mathrm{exp}[-\Gamma(E_{R}/E_{th}-1)]$
is chosen, with $\Gamma=4.3\pm0.3$ and the threshold energy $E_{th}=8$
keV.

For COUPP~\cite{Behnke:2012ys}, we adopt a similar exponential efficiency
$\eta_{\mathrm{C,F}}=1-\mathrm{exp}[-\alpha(E_{R}/E_{th}-1)]$ with
$\alpha=0.15$ for fluorine and carbon, and $\eta_{\mathrm{I}}=1$
for iodine above the nucleation threshold. We obtain 90\% C.L. exclusion contours
 by Poisson statistics with 13 observed events against
the expected 4.5 background events~\cite{Cannoni:2012jq}. Here we
note that for the COUPP experiment we adopt the exponential bubble
nucleation efficiency, which results in a more conservative exclusion
limit when compared to the flat model.

For Xenon100, we follow \cite{Aprile:2011hx,Aprile:2013doa} to derive
the 90\% C.L. limit curves by using the maximum gap method~\cite{Yellin:2002xd},
under the assumption that the expected S1 signals are subject to the
Poisson fluctuation, and taking into account the finite photomultiplier
resolution $\sigma_{\mathrm{PMT}}=0.5~\mathrm{\, PE}$(photon electron).
Throughout our study the dark matter is assumed to be distributed
in an isothermal halo with a local density $\rho_{\chi}=0.3~\mathrm{GeV/cm^{3}}$
and a Maxwellian velocity distribution with a dispersion $v_{0}=220~\mathrm{km/s},$
truncated at the galactic escape velocity $v_{\mathrm{esc}}=544~\mathrm{km/s}$.
We present the constraints for a Dirac fermionic WIMP in Fig. \ref{fig:90=000025-C.L.-exclusion pn}
and Fig. \ref{fig:90=000025-C.L.-exclusion n,p-1}

In Fig. \ref{fig:90=000025-C.L.-exclusion pn} we show the 90\% C.L.
exclusion limits on $\sigma$ for the normalized operators listed
in Table \ref{tab:dark-matter form factor} for elastic scattering in the case of $a_{p}=a_{n}$.
SIMPLE and Xenon100 are the most sensitive in the low and high WIMP
mass ranges, respectively. The 90\% C.L. exclusion contours on $\sigma_{n}$
($a_{p}=0$) and $\sigma_{p}$($a_{n}=0$) are shown in Fig. \ref{fig:90=000025-C.L.-exclusion n,p-1}.
As discussed in Ref. \cite{Fitzpatrick:2012ix,Fitzpatrick:2012ib},
the nuclear response of operator $\mathcal{O}_{3}$ ($\mathcal{O}_{3}/q^{2}$)
may tend to favor a heavy element and hence bear a similarity to
the standard SI response. One notes that the limit contour of operator
$\mathcal{O}_{3}$ ($\mathcal{O}_{3}/q^{2}$) features a sharp decline
around 10 GeV, where the response of the heavy element iodine is suddenly
switched on due to the step function of the  nucleation efficiency.

\section{\noun{dark matter capture by sun}}

\subsection{Capture rate and annihilation rate}

The capture of dark matter takes place when the incoming WIMPs collide
with the solar elements and are stripped of enough kinetic energy
to escape from the gravitational pull of the Sun. We calculate this
process in the Sun's rest frame, in which the WIMP velocity distribution
is written as

\begin{equation}
f(\mathbf{u})=\frac{e^{-\frac{(\mathbf{u}+\mathbf{v}_{\odot})^{2}}{v_{0}^{2}}}}{N(v_{\mathrm{esc}})},
\end{equation}
where $\mathbf{v}_{\odot}$ is the velocity of the Sun and $\mathbf{u}$
is the DM velocity at infinity with respect to the Sun's rest frame,
$N(v_{\mathrm{esc}})$ being the normalization constant dependent
on $v_{\mathrm{esc}}$. Due to the smallness of the WIMP-nucleus cross
section, the Sun is assumed to be optically thin to the incoming WIMPs
and hence multiple scatterings are neglected. By use of the DM angular momentum conservation
in the solar central field, one can obtain the following WIMP scattering
event rate $R_{\odot}$~\cite{Gould:1987ir},

\begin{equation}
R_{\odot}=\sum_{A_{i}}\int_{\mathrm{Sun}}dV\int\frac{f(\mathbf{u})}{u}w\Omega_{A_{i}}^{-}(w)d^{3}u,\label{eq:capture rate 1}
\end{equation}
in which the summation is taken over all elements $\{A_{i}\}$ in
the Sun. $\Omega_{A_{i}}^{-}(w)$ is defined as

\begin{equation}
\Omega_{A_{i}}^{-}(w)=n_{A_{i}}(r)n_{\chi}\sigma(w)w.\label{eq:omega minus}
\end{equation}

$n_{A_{i}}(r)$ is the number density of element $A_{i}$ at radius
$r$ and the local WIMP density $n_{\chi}$ in the solar neighborhood
is determined by $n_{\chi}=\rho_{\chi}/m_{\chi}$. $w(r)=\sqrt{u^{2}+v_{\mathrm{esc}}^{2}(r)}$
is the incident DM velocity at radius $r$ inside the Sun, accelerated
from the initial velocity $u$ at infinity by the solar gravitational
attraction. $v_{\mathrm{esc}}(r)$ , the escape velocity at radius
$r$, is related to that at the Sun's center $v_{c}=1354~\mathrm{km/s}$
and at surface $v_{s}=795~\mathrm{km/s}$ by the following approximate
relation~\cite{Gould:1991hx}:

\begin{equation}
v_{\mathrm{esc}}^{2}(r)=v_{c}^{2}-\frac{M(r)}{M_{\odot}}(v_{c}^{2}-v_{s}^{2}).
\end{equation}

$M_{\odot}$ is the mass of the Sun and $M(r)$ is the mass contained
within radius $r$. Since only those WIMPs that lose enough energy
after scattering can be trapped by the Sun, for
the solar capture it is then demanded  that the scattered WIMPs be contained in the radius
of Jupiter's orbit $r_{0}$~\cite{Kumar:2012uh,Peter:2009mk}, which
implies

\begin{equation}
\frac{q^{2}}{2m_{A}}\geq\frac{m_{\chi}u^{2}}{2}+\frac{m_{\chi}v_{\mathrm{esc}}^{2}(r_{0})}{2},\label{eq:q-mini}
\end{equation}
with $v_{\mathrm{esc}}(r_{0})=18.5~$km/s. Therefore, to derive the
capture rate $C_{\odot}$, one can replace $\sigma(w)$ in Eq. (\ref{eq:omega minus})
with an effective capture cross section,

\begin{equation}
\sigma_{\mathrm{SD},A_{i}}^{\mathrm{cap}}(w)=\frac{16}{3}\frac{\sigma}{2\mu_{p}^{2}w^{2}}\int_{q_{\mathrm{min}}}^{2\mu_{A_{i}}w}F_{\chi-T}^{2}(q,w)qdq,\label{eq:effective capture cross section}
\end{equation}
where $q_{\mathrm{min}}$ is determined from Eq.~(\ref{eq:q-mini}).
Then we have

\begin{equation}
C_{\odot}=\sum_{A_{i}}\int_{\mathrm{Sun}}dV\int\frac{f(\mathbf{u})}{u}w\Omega_{\mathrm{cap},A_{i}}^{-}(w)d^{3}u,\label{eq:capture rate2}
\end{equation}
with

\begin{equation}
\Omega_{cap,A_{i}}^{-}=n_{A_{i}}(r)n_{\chi}\sigma_{\mathrm{SD},A_{i}}^{\mathrm{cap}}(w)w.
\end{equation}

One should note that besides depicting the motion of the WIMPs in
a more realistic three-body interaction picture, the introduction of such
a finite radius $r_{0}$ simultaneously avoids the divergence disaster
that we encounter in the calculation of capture rate in the massless
mediator scenario. So it is reasonable to expect that a massless force
carrier will give a good representative description of one with a
low but finite mass for capture.

Moreover, if we make the approximation by setting $v_{\mathrm{esc}}\rightarrow\infty$,
the expression of differential capture rate can be simplified remarkably
to

\begin{equation}
\frac{dC_{\odot}}{dV}=\int_{0}^{\infty}\frac{4\pi u^{2}du}{(\pi v_{0}^{2})^{3/2}}\frac{v_{0}^{2}}{2uv_{\odot}}\Omega_{\mathrm{cap},A_{i}}^{-}(w)\frac{w}{u}\exp[-\left(\frac{u^{2}+v_{\odot}^{2}}{v_{0}^{2}}\right)]\sinh\left(\frac{2uv_{\odot}}{v_{0}^{2}}\right).
\end{equation}

As we merely focus on the spin-dependent operators in this paper,
only contribution of hydrogen atoms to the capture rate is relevant
in our calculation. This is because all the SD operators in Table\ref{tab:dark-matter form factor}
except $\mathcal{O}_{3}$ and $\mathcal{O}_{3}/q^{2}$ lead to amplitudes
proportional to the spin or angular momentum of the nucleus at small
$q$ , rather than the square of atomic number $A_{i}$. It indicates
that the contribution from heavy elements are subject to a significant
abundance-suppression. As for $\mathcal{O}_{3}$ and $\mathcal{O}_{3}/q^{2}$,
however, the amplitudes may grow with atomic number~\cite{Fitzpatrick:2012ix},
it then requires a thorough understanding about the relevant nuclear
structure of those heavy elements present in the Sun, which is beyond
the scope of this work. The distributions of the elements are obtained
from the Standard Sun Model (SSM) GS98~\cite{Serenelli:2009yc}.

Under the assumption of a large WIMP mean free path, the trapped WIMPs
thermalize and sink into the core of the Sun, so
the WIMP annihilation takes place at the center region , depleting
the WIMP population through annihilation and evaporation. The evolution
of the WIMP number $N$ in the Sun is described by the following equation,

\begin{equation}
\overset{\cdot}{N}=C_{\odot}-A_{\odot}N^{2}-E_{\odot}N,\label{eq:N equation}
\end{equation}
which includes the effects of capture ($C_{\odot}$), annihilation
($A_{\odot}$), and evaporation ($E_{\odot}$). The annihilation rate
$A_{\odot}$ is defined as

\begin{equation}
A_{\odot}=\frac{\langle\sigma v\rangle_{\odot}}{V_{\mathrm{eff}}},
\end{equation}
where $\langle\sigma v\rangle_{\odot}$ is the thermal average over
the annihilation cross section times the relative velocity and $V_{\mathrm{eff}}$
is the effective volume for annihilation, which can be approximately
given as~\cite{Griest:1986yu,Gould:1987ir}

\begin{equation}
V_{\mathrm{eff}}=5.8\times10^{30}\mathrm{cm^{3}}\left(\frac{1\mathrm{GeV}}{m_{\chi}}\right)^{3/2}.
\end{equation}

The evaporation mass $m_{ev}$ is defined as a characteristic parameter
above which the WIMP evaporation effect is negligible. As an estimate
the evaporation mass for a SD cross section $\sigma_{\mathrm{SD}}\sim4\times10^{-36}\mathrm{\, cm^{2}}$
is about $3\,\mathrm{GeV}$ \cite{Spergel:1984re}, so we neglect
the evaporation effect in the WIMP mass range of our interest ($m_{\chi}\geqslant5\,\mathrm{GeV}$),
considering $m_{ev}$ depends on the WIMP-hydrogen cross
section in a logarithmic manner for the rare scattering scenario \cite{Spergel:1984re,Griest:1986yu,Gould:1987ju}. Thus, one
can easily obtain the solution to Eq.(\ref{eq:N equation}) as

\begin{equation}
N(t)=\sqrt{\frac{C_{\odot}}{A_{\odot}}}\tanh(\sqrt{C_{\odot}A_{\odot}}\, t),
\end{equation}
so the present annihilation rate can be immediately written as

\begin{equation}
\Gamma_{\odot}=\frac{1}{2}A_{\odot}N^{2}(t_{\odot})=\frac{1}{2}C_{\odot}\tanh^{2}(\sqrt{C_{\odot}A_{\odot}}\, t_{\odot}).\label{eq:gamma sun}
\end{equation}

$t_{\odot}\backsimeq4.5\times10^{9}$ yr is the age of the Sun. If
$\sqrt{C_{\odot}A_{\odot}}\, t_{\odot}\gg1$, the DM capture-annihilation
process reaches equilibrium, and as a result the annihilation rate
is solely determined by capture rate through $\Gamma_{\odot}=\frac{1}{2}C_{\odot}.$

\subsection{Constraints from Super-Kamionkande and IceCube}

The differential flux of muon neutrino observed at the Earth for annihilation
channel $f$ is

\begin{equation}
\frac{d\Phi_{\nu_{\mu}}^{f}}{dE_{\nu_{\mu}}}=\frac{\Gamma_{\odot}}{4\pi d^{2}}\frac{dN_{\nu_{\mu}}^{f}}{dE_{\nu_{\mu}}},
\end{equation}
where $d$ is the Earth-Sun distance, and $dN_{\nu_{\mu}}^{f}/dE_{\nu_{\mu}}$
is the differential energy spectrum of the muon neutrino. In order
to fully and accurately determine the neutrino spectrum, a wide variety
of phenomena must be taken into consideration including the hadronization
of quarks, neutrino oscillations, energy loss in the solar medium
and en route to the Earth, etc. Here we choose the following neutrino
oscillation parameters~\cite{An:2012eh,Tortola:2012te}:

\[
\sin^{2}\theta_{12}=0.32,\quad\sin^{2}\theta_{23}=0.49,\quad\sin^{2}\theta_{13}=0.026,\quad\delta=0.83\pi,
\]

\begin{equation}
\Delta m_{21}^{2}=7.62\times10^{-5}\mathrm{eV^{2}},\quad\Delta m_{31}^{2}=2.53\times10^{-3}\mathrm{eV^{2}}.
\end{equation}

The upgoing muons produced from the interactions between the arriving
neutrinos and the Earth rocks or ice can be detected by the water
Cherenkov detector Super-Kamionkande~\cite{Tanaka:2011uf} and the
neutrino telescope IceCube~\cite{Abbasi2012,Aartsen:2012kia}, then
one can map the upper limits on the muon flux into the constraints
on the WIMP annihilation rate by assuming specific annihilation modes
and considering relevant details in neutrino propagation.

\begin{figure}[H]
\begin{centering}
\includegraphics[scale=0.45]{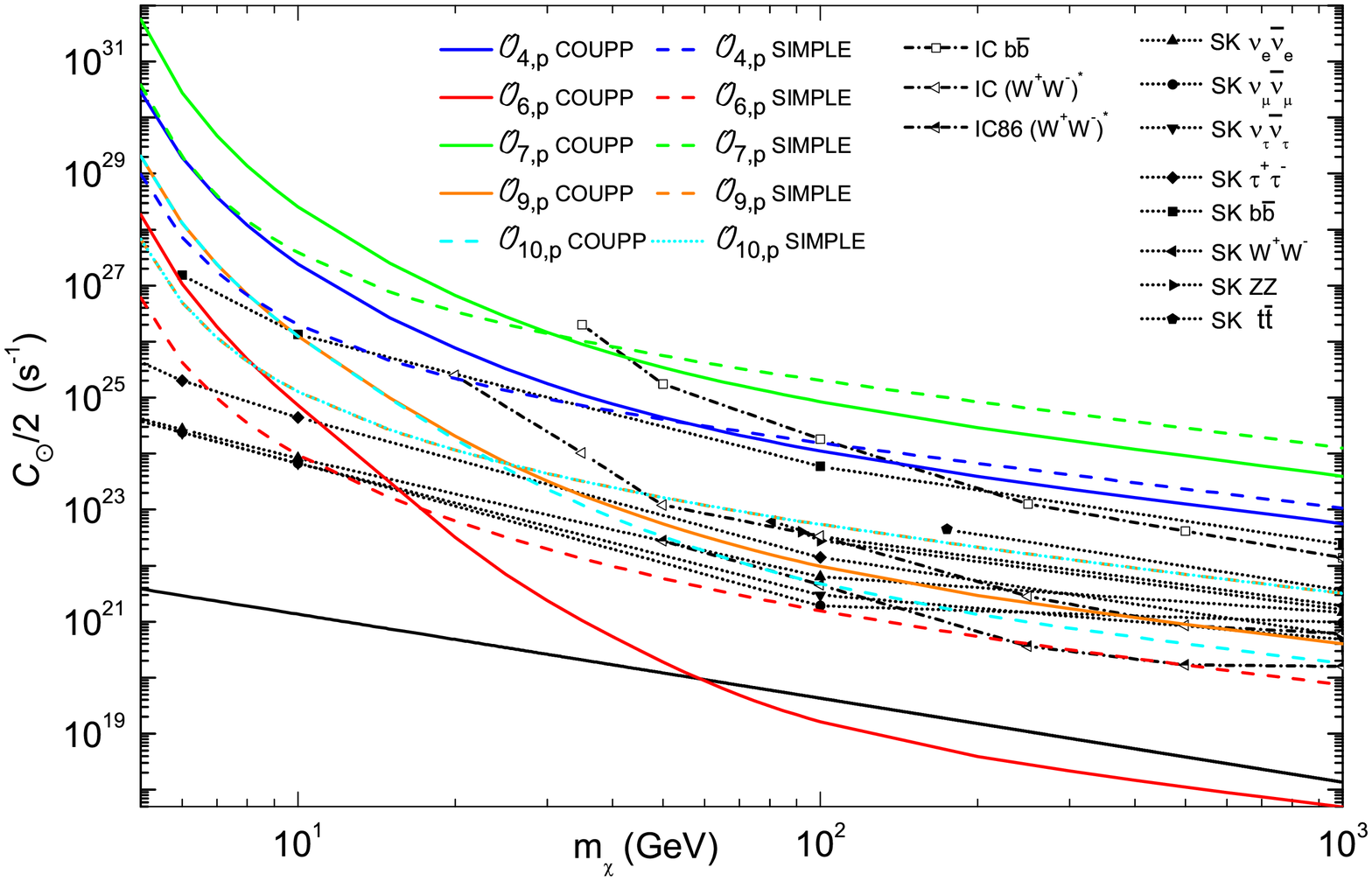}
\par\end{centering}

\begin{centering}
\includegraphics[scale=0.45]{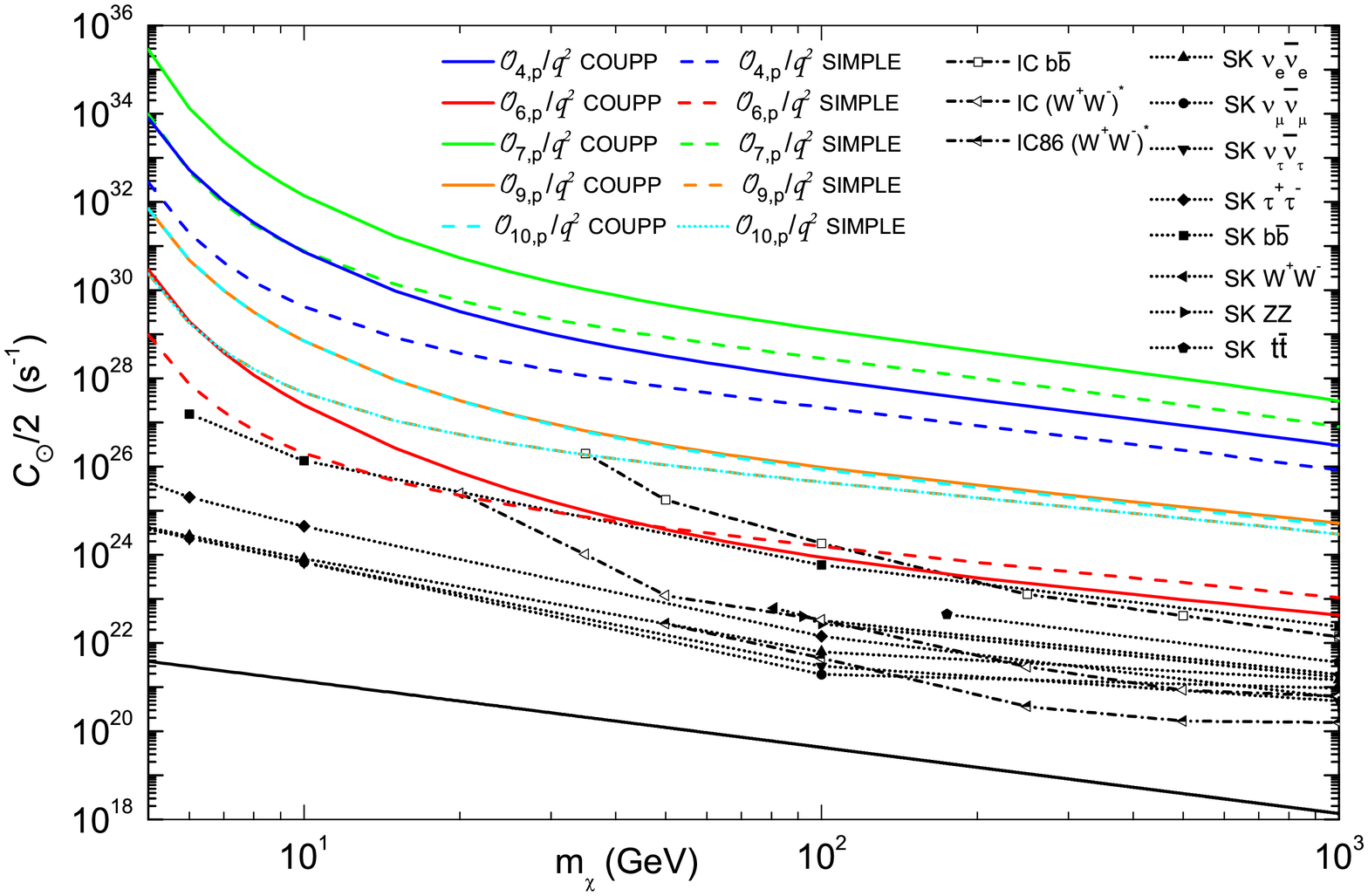}
\par\end{centering}

\caption{the 90\% C.L. upper limits on $C_{\odot}/2$ for various Dirac fermionic
WIMP SD operators for elastic scattering in Table \ref{tab:dark-matter form factor} in the
case $a_{n}=0$, under the assumption that the equilibrium between
the WIMP capture and annihilation is reached. The black line denotes
the equilibrium criterion detailed in the text. To clearly illustrate
the results, the constraints are shown in two separate panels.\label{fig:bounds on capture-p}}
\end{figure}

\begin{figure}[H]
\noindent \begin{centering}
\includegraphics[scale=0.45]{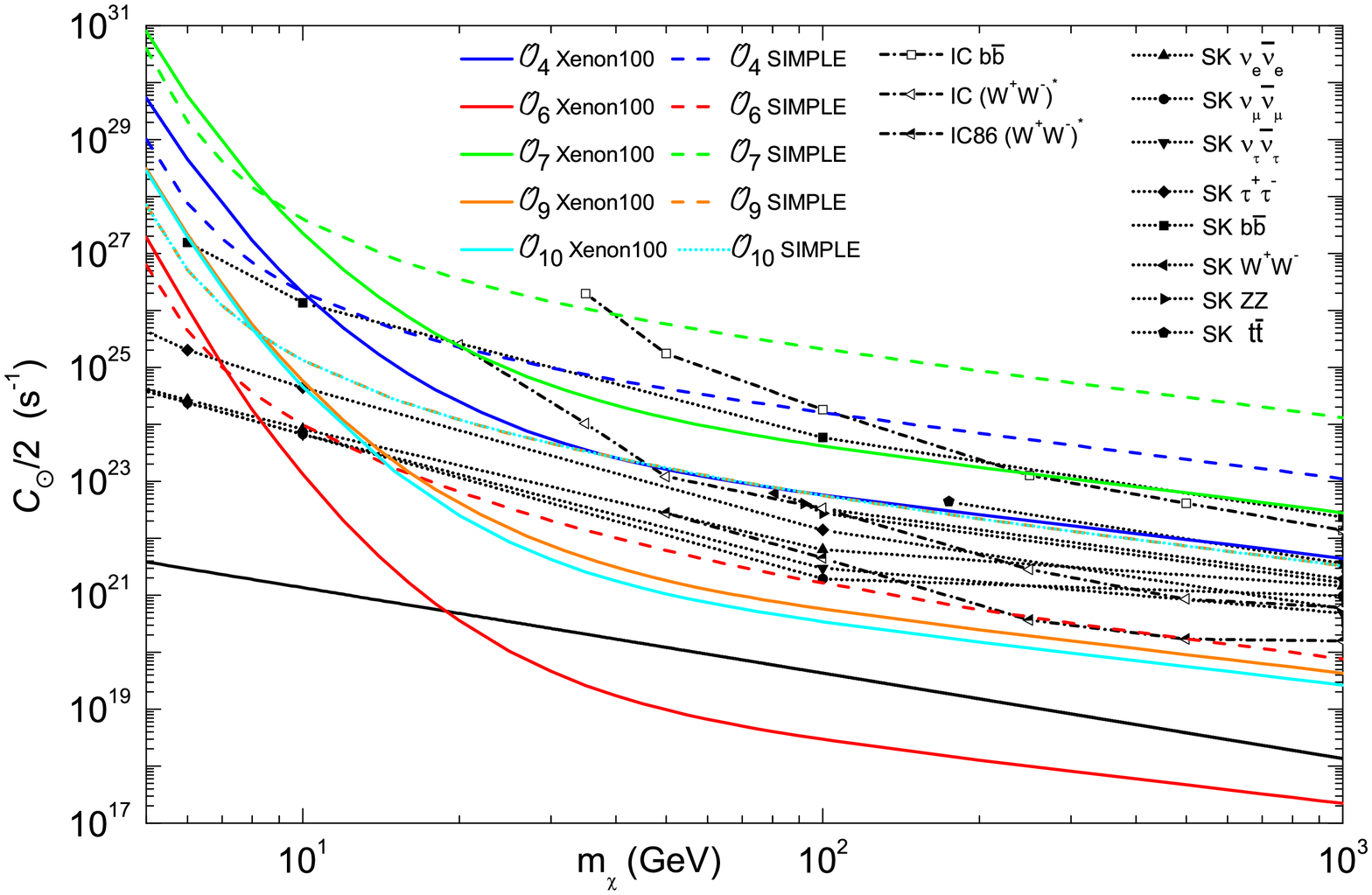}
\par\end{centering}

\noindent \begin{centering}
\includegraphics[scale=0.45]{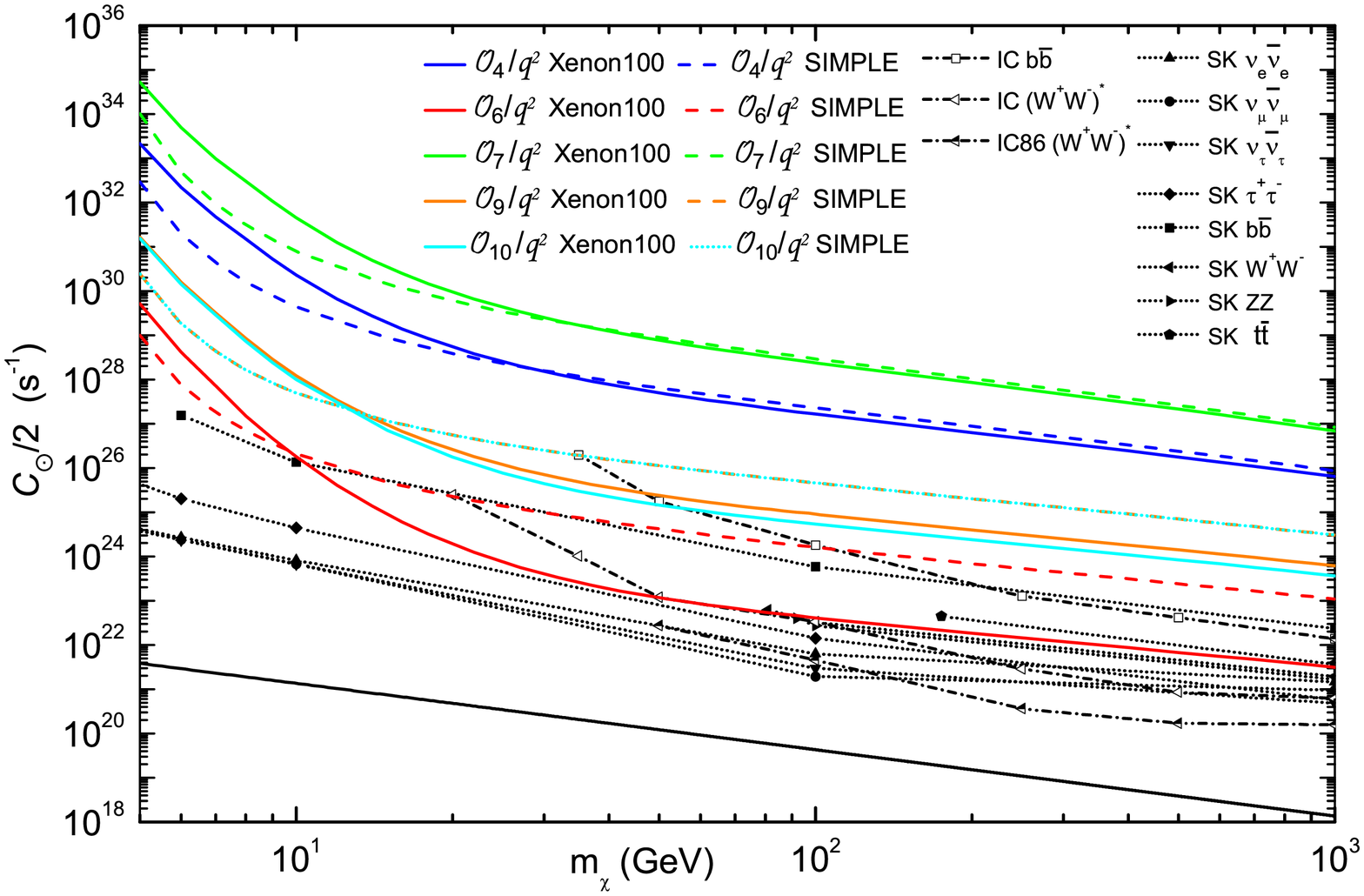}
\par\end{centering}

\caption{the 90\% C.L. upper limits on $C_{\odot}/2$ for various SD operators for elastic scattering
in Table \ref{tab:dark-matter form factor} in the case $a_{n}=a_{p}$,
under the assumption that the equilibrium between the DM capture and
annihilation is reached. The black line denotes the equilibrium criterion
detailed in the text. To clearly illustrate the results, the constraints
are shown in two separate panels. \label{fig:bound on capture-pn}}
\end{figure}

For Super-Kamionkande the neutrino-induced muon events are divided
into three categories: fully-contained, stopping and through-going,
and the fraction of each category as a function of the parent neutrino
energy $E_{\nu_{\mu}}$ is shown in Fig. 2 in Ref.~\cite{Tanaka:2011uf}.
On the other hand, the IceCube Collaboration has also reported the
constraints on the DM annihilation rate $\Gamma_{\odot}$ for $\overline{b}b$
, $W^{+}W^{-}$, and $\tau^{+}\tau^{-}$($m_{\chi}$< 80.4 GeV)channels
in Table  \uppercase\expandafter{\romannumeral1} of Ref.~\cite{Aartsen:2012kia},
together with the expected 180-day sensitivity of the completed IceCube
detector . Here we adopt the relevant detection parameters summarized
in Table \uppercase\expandafter{\romannumeral2} and the 90\% C. L.
upper limits on the annihilation rates from Ref.~\cite{Guo:2013ypa}.
In order to show how the equilibrium assumption remains a good approximation,
we take the criterion that $\sqrt{C_{\odot}A_{\odot}}\, t_{\odot}\geq3.0$
($\tanh^{2}(\sqrt{C_{\odot}A_{\odot}}\, t_{\odot})\geq0.99$) or $C_{\odot}/2\geq4.3\times10^{22}(1\mathrm{GeV}/m_{\chi})^{3/2}\mathrm{s}^{-1}$
for the $s$-wave thermally averaged annihilation cross section $\langle\sigma v\rangle_{\odot}\approx3.0\times10^{-26}~\mathrm{\mathrm{cm^{3}}s^{-1}}$,
which is plotted in a black solid line in Fig. \ref{fig:bounds on capture-p}
and Fig. \ref{fig:bound on capture-pn} . We can see that the constraints
deduced from the neutrino detectors are far above the equilibrium
contour, which assures the equilibrium assumption $\Gamma_{\odot}=\frac{1}{2}C_{\odot}$.

Meanwhile we translate the above-acquired constraints on the elastic
WIMP-proton coupling $a_{p}$ into the bounds on the annihilation
rate $\Gamma_{\odot}$ by involving only hydrogen presence in our
calculation. All these results are shown in Fig. \ref{fig:bounds on capture-p}
and Fig. \ref{fig:bound on capture-pn}. To project the upper limits
on the elastic WIMP-nucleon couplings onto the annihilation rate, we only
need to replace the parameter $\sigma$ and the WIMP-nucleus form factor
$F_{\chi-T}^{2}$ in Eq.(\ref{eq:effective capture cross section})
with the corresponding upper limit $\sigma^{\mathrm{limit}}(a^{\mathrm{limit}})$
drawn from direct detections and the WIMP-hydrogen form factor. For
instance, the constraint on the effective capture cross section for
operator $\mathcal{O}_{9,p}=i\mathbf{S}_{\chi}\cdot(\mathbf{S}_{p}\times\mathbf{q})$
can be obtained as the following,

\begin{eqnarray}
\sigma_{\mathrm{SD},\mathcal{O}_{9},p}^{\mathrm{cap,limit}}(w) & = & \frac{(a_{p}^{\mathrm{limit}})^{2}}{2\pi w^{2}}\frac{1}{(2s_{p}+1)(2s_{\chi}+1)}\sum_{\mathrm{spins}}\int_{q_{\mathrm{min}}}^{2\mu_{\mathrm{p}}w}|\langle i\mathbf{S}_{\chi}\cdot(\mathbf{S}_{p}\times\mathbf{q})\rangle|^{2}qdq\nonumber \\
 & = & \frac{2}{3}\frac{\sigma_{\mathcal{O}_{9},p}^{\mathrm{limit}}}{2\mu_{p}^{2}w^{2}}\int_{q_{\mathrm{min}}}^{2\mu_{\mathrm{p}}w}\left(\frac{q}{q_{0}}\right)^{2}qdq,\label{eq:capture cross section-9}
\end{eqnarray}where $\mathbf{S}_{p}$($s_{p}$) represents the spin of the hydrogen
atom (proton).

\vspace{1.5cm}

We can also derive the upper limit on the effective
capture cross section for operator $\mathcal{O}_{10,p}=i\mathbf{S}_{p}\cdot\mathbf{q}$
in a similar way,

\begin{equation}
\sigma_{\mathrm{SD},\mathcal{O}_{10},p}^{\mathrm{cap,limit}}(w)=\frac{4}{3}\frac{\sigma_{\mathcal{O}_{10},p}^{\mathrm{limit}}}{2\mu_{p}^{2}w^{2}}\int_{q_{\mathrm{min}}}^{2\mu_{\mathrm{p}}w}\left(\frac{q}{q_{0}}\right)^{2}qdq.\label{eq:capture cross section-10}
\end{equation}

It is worth noting that unlike the case of direct detection, in which
$\sigma$ depends on the normalization parameters $q_{0}$ and $v_{0}$,
the constraints on the annihilation rate (or capture rate) merely
depend on the operators in the first column in Table \ref{tab:dark-matter form factor},
as shown in the first line of Eq.~(\ref{eq:capture cross section-9}).

One can learn from Fig. \ref{fig:bounds on capture-p} that when the
proton-coupling dominant scenario($a_{n}=0$) is assumed , the Super-Kamionkande
provides more stringent constraints on SD operators $\mathcal{O}_{4}$
and $\mathcal{O}_{7}$ than the direct detection experiments SIMPLE
and COUPP do. However, for $\mathcal{O}_{6}$ both SIMPLE and COUPP
become more sensitive for the DM mass range $m_{\chi}>20$~GeV. For
the light mediator case, two neutrino detectors are proved to be more
effective in exploring the parameter space. Similar situations can
be found in Fig.~\ref{fig:bound on capture-pn} for the equal coupling
scenario($a_{p}=a_{n}$). On the other hand, for the simplest SD interaction
$\mathcal{O}_{4}$ with only WIMP-proton coupling shown in the top
panel in Fig. \ref{fig:bounds on capture-p}, the neutrino-based constraints
lead in the detection sensitivity over that of the direct detection
approach by a factor up to $2\sim3$ orders of magnitude, depending
on the specific annihilation channels. However, with the power of
the transferred momentum $q$ increased in the effective operator,
the direct detections turn more effective in excluding the parameter
space, especially in the large WIMP mass region that favors a large transferred
momentum. For $\mathcal{O}_{6}\propto q^{2}$, both constraints from
SIMPLE and COUPP reach below that of the indirect search Super-Kamionkande
and IceCube in the region $m_{\chi}>20$ GeV. For the same reason,
considering a propagator inversely proportional to $q^{2}$, one would
expect that the bounds tend to go upwards in the massless mediator scenario
in the bottom panel of Fig. \ref{fig:bounds on capture-p}. Similar
arguments still hold for the equal coupling case, where the stringent
constraints on the couplings in direct detection experiments can also
be obtained by Xenon100, mainly through the WIMP-neutron interaction.

It is noted that the constraints on the annihilation rates of $\mathcal{O}_{9,p}$
and $\mathcal{O}_{10,p}$ are coincident with each other in Fig.~\ref{fig:bounds on capture-p}.
To explain this coincidence, we first compare the two upper limits
$\sigma_{p,9}^{\mathrm{limit}}$ and $\sigma_{p,10}^{\mathrm{limit}}$
from Table \ref{tab:dark-matter form factor} and find the following
relation:

\begin{equation}
\frac{\sigma_{\mathcal{O}_{10},p}^{\mathrm{limit}}}{\sigma_{\mathcal{O}_{9},p}^{\mathrm{limit}}}=\frac{1}{4}\frac{F_{\Sigma^{'}}^{(p,p)}}{F_{\Sigma^{''}}^{(p,p)}}.
\end{equation}
 So in light of Eq.~(\ref{eq:capture cross section-9}) and Eq.~(\ref{eq:capture cross section-10})
we further have

\begin{equation}
\frac{C_{\odot,\mathcal{O}_{10},p}^{\mathrm{limit}}}{C_{\odot,\mathcal{\mathcal{O}}_{9},p}^{\mathrm{limit}}}=\frac{1}{2}\frac{F_{\Sigma^{'}}^{(p,p)}}{F_{\Sigma^{''}}^{(p,p)}},
\end{equation}
where $C_{\odot,\mathcal{\mathcal{O}}_{9},p}^{\mathrm{limit}}$ and
$C_{\odot,\mathcal{O}_{10},p}^{\mathrm{limit}}$ are the relevant
upper limits on the capture rate of $\mathcal{O}_{9,p}$ and $\mathcal{O}_{10,p}$,
respectively. Since the ratio between the transverse form factor
$F_{\Sigma^{'}}^{(p,p)}$ and the  longitudinal one $F_{\Sigma^{''}}^{(p,p)}$
approaches  $2$ in the long-wavelength limit and fluorine favors
a small transferred momentum in direct detection, one can understand
why the constraints on the annihilation rates of $\mathcal{O}_{9,p}$ and
$\mathcal{O}_{10,p}$ are inseparable in Fig. \ref{fig:bounds on capture-p}.
Similar arguments can be applied to explain the coincidence of $\mathcal{O}_{9}$
and $\mathcal{O}_{10}$ in Fig. \ref{fig:bound on capture-pn}.

\section{\noun{discussion and conclusion}}

In this paper we have studied how a diversity of effective SD operators
can lead to different interpretations of some direct detection experimental
results, based on the nuclear form factors given in \cite{Fitzpatrick:2012ix}.
For each type of operator, we further group the possible interactions
into three illustrative categories: $a_{n}=0$, $a_{p}=0,$ and $a_{p}=a_{n}$.
When the WIMP couples dominantly with proton (neutron) over neutron
(proton), the first (second) category lives up to a good approximation,
and when the two coupling strengths are comparable, we  expect
the third category $a_{p}=a_{n}$ to give a representative description.
We have used proton-sensitive experiments SIMPLE and COUPP to plot
the upper limits on the proton coupling ($\sigma_{p}$) and used neutron-sensitive
experiment Xenon100 to constrain the neutron coupling ($\sigma_{n}$).
One can draw from Fig.~\ref{fig:90=000025-C.L.-exclusion n,p-1} that
SIMPLE is more effective in excluding parameter space below the WIMP
mass of tens of GeV, whereas COUPP turns more strict in the larger
WIMP mass range due to the heavy element iodine that favors a larger
recoil energy. However, when the massless mediator scenario is involved
in consideration, the propagator provides an enhancement in the low
transferred momentum regime which makes SIMPLE more sensitive in the
whole WIMP mass range. In Fig.~\ref{fig:90=000025-C.L.-exclusion pn}
we have shown the 90\% C. L. upper limits on the SD operators for
the equal coupling scenario, in which one can see the Xenon100 and
SIMPLE give  complementary constraints on $\sigma$ in combination.

Since all those SD operators (except $\mathcal{O}_{3}$ and $\mathcal{O}_{3}/q^{2}$)
listed in Table \ref{tab:dark-matter form factor} result in scattering
amplitudes proportional to the angular momentum or spin of the solar
elements in the long-wavelength limit, unlike the SI amplitudes that
are usually with an $A^{2}$ enhancement, we have only calculated
the contribution of hydrogen to the WIMP capture rate. This is because
other solar component elements related to the SD interactions are
significantly suppressed due to their low abundance in the Sun. As
for operator $\mathcal{O}_{3}$ (or $\mathcal{O}_{3}/q^{2}$), the
nuclear response is sensitive to $\sum_{i}\mathbf{L}_{i}\cdot\mathbf{S}_{i}$
which may favor heavier elements, so that we must take heavy elements
into our consideration just as  we did in the SI case. Unfortunately,
so far there is no such knowledge of the nuclear form factors of the
relevant elements, so we should leave this investigation to future
work. We have also mapped the constraints on the WIMP-nucleon couplings
onto the bounds on the annihilation rate deduced from the solar neutrino
experiments, Super-Kamionkande and IceCube, under the equilibrium
assumption, which allows us to provide  complementary exclusion contours
from both direct and indirect detection experiments.

Finally, we point out that our discussions concerning the WIMP capture
and distribution are based on the assumption of a large Knudsen number
($K>1$) \cite{Spergel:1984re,Gould:1989ez}. In that picture, the
WIMP's free path is much larger than the length of the WIMP populated
region, and its distribution can be described as isothermal with a
characteristic temperature $T_{\chi}$ \cite{Spergel:1984re}. Strictly speaking,  the large Knudsen number assumption may no longer be valid for
some nonconventional operators at the parameter ($\sigma$) scales probed
by the present direct detection experiments. For instance, in the
case where the WIMP and nucleus interact through a light force carrier,
the relevant cross section for WIMPs that reside in the Sun's core
will receive a boost factor relative to the one required for capture.
If the boost factor is so large that the WIMPs and nuclei collide
frequently in the inner part of the Sun, the interaction moves into
the local thermal equilibrium region (LTE) \cite{Gould:1989hm}, in
which the WIMP distribution and energy transport are discussed recently by
other authors in Ref. \cite{Vincent:2013lua}. To study the solar DM signals for some unconventional operators in a more realistic way, we have to take the LTE scenario into consideration, which we leave to our future work.
\begin{acknowledgments}
We thank Wanlei Guo for useful discussions. This work is supported
in part by the National Basic Research Program of China (973 Program)
under Grant No. 2010CB833000, the National Nature Science Foundation
of China (NSFC) under Grants No. 10821504 and No. 10905084, and the
Project of Knowledge Innovation Program (PKIP) of the Chinese Academy
of Science.
\end{acknowledgments}
%\bibliographystyle{JHEP}
%\bibliography{ref1}

\providecommand{\href}[2]{#2}\begingroup\raggedright\endgroup

\end{document}